\documentclass[pra,amsmath,amssymb, twocolumn, showpacs]{revtex4}

\usepackage{dcolumn}

\usepackage{bm}

\usepackage[dvips]{graphicx}

\usepackage{epsfig}

\begin{document}

\title{Complete time-dependent treatment of a three-level system}

\author{A.\  R.\  P. Rau$^{*}$ and Weichang Zhao}
\affiliation{Department of Physics and Astronomy, Louisiana State University,
Baton Rouge, Louisiana 70803-4001}


\begin{abstract}

Both unitary evolution and the effects of dissipation and decoherence for a general three-level system are of wide interest in quantum optics, molecular physics, and elsewhere. A previous paper presented a technique for solving the time-dependent operator equations involved but under certain restrictive conditions. We now extend our results to a general three-level system with arbitrary time-dependent Hamiltonians and Lindblad operators. Analytical handling of the SU(3) algebra of the eight operators involved leaves behind a set of coupled first-order differential equations for classical functions. Solution of this set gives a complete solution of the quantum problem, without having to invoke rotating-wave or other approximations. Numerical illustrations for multiphoton couplings and quantum control are given.
\end{abstract}

\pacs{03.65.Yz, 05.30.-d, 42.50.Lc, 32.80.Qk}

\maketitle

\section{Introduction}

Three-level systems arise in many physical contexts. A spin-1 particle in an external magnetic field, three states of atoms or molecules coupled by lasers or other interactions, and oscillations among the three neutrino flavors are just a few varied examples that lead to a Schr\"{o}dinger equation for the wave function or, equivalently, for the evolution operator $U(t)$,

\begin{equation}
idU(t)/dt = H(t)U(t), \,\, U(0)= \mathcal{I}.
\label{eqn1}
\end{equation}
Here, and throughout, we set $\hbar =1$. In a matrix representation, $H(t)$ is a $3 \times 3$ matrix. As examples, population trapping and dispersion was considered through \cite{ref1}

\begin{equation}
H=\left( 
\begin{array}{ccc}
0 & G_{1}^{*} & 0 \\ 
G_1 & \Delta_1 & G_{2}^{*} \\ 
0 & G_2 & \Delta_1 + \Delta_2
\end{array}
\right),
\label{eqn2}
\end{equation}
where the coefficients $G$ represent couplings between level one and two and between two and three while the diagonal elements give the energy levels and detunings. Multiphoton coupling in molecular systems has used \cite{ref2}

\begin{equation}
H=\left( 
\begin{array}{ccc}
-E_1 & \Omega_{12} (t) & 0 \\ 
\Omega_{12} (t) & 0 & \Omega_{23} (t) \\ 
0 & \Omega_{23}(t) & \delta
\end{array}
\right),
\label{eqn3}
\end{equation}
while recent treatments of quantum control in the rotating-wave approximation considered \cite{ref3,ref4}

\begin{equation}
H=\left( 
\begin{array}{ccc}
0 & \Omega_{12} (t) & 0 \\ 
\Omega_{12} (t) & \Delta & \Omega_{23} (t) \\ 
0 & \Omega_{23}(t) & 0
\end{array}
\right).
\label{eqn4}
\end{equation}
With constant coefficients for the diagonal entries, the two Hamiltonians in Eq.~(\ref{eqn3}) and Eq.~(\ref{eqn4}) are essentially equivalent upon shifting the zero of the energy scale and suitable identification of the constants.

Some of these studies introduced decay phenomenologically by considering complex values for the diagonal entries. A fuller treatment of dissipation and decoherence proceeds, however, by replacing the unitary evolution equation in Eq.~(\ref{eqn1}) by master equations for the density matrix $\rho $. A widely used class of such equations is the Liouville-von Neumann-Lindblad equation \cite{ref5, ref6},

\begin{eqnarray}
i\dot{\rho} & = & [H,\rho ]+
\frac{1}{2}i\!\sum_{k}\left( [L_k\rho,L_k^{\dagger }]+
[L_k,\rho L_k^{\dagger }]\right)  \nonumber \\
& = & [H,\rho ]-\frac{1}{2}i\!\sum_{k}\left( L_k^{\dagger }L_k\rho +\rho
L_k^{\dagger }L_k-2L_k\rho L_k^{\dagger }\right)\!, \label{eqn5}
\end{eqnarray}

\noindent where an over-dot denotes differentiation with respect to time, and the $L_k$ are operators in the system through which dissipation and decoherence are introduced. Even though
this can result in non-unitary evolution, the form of the equation preserves Tr($\rho$) and
positivity of probabilities. A more mathematical discussion of such ``super-operators" and
 ``dynamical semigroups" is given in \cite{ref7}. Note that preservation of the trace so that there is no overall dissipation hinges on the coefficient 2 in the last term above, any other value leading to a change in the value of the trace with time.
 
In recent papers \cite{ref6,ref8}, we have developed a technique for solving Eq.~(\ref{eqn1}) and Eq.~(\ref{eqn5}). This method, which seems to have been independently rediscovered several times over the decades, with the earliest reference going at least back to Wei and Norman \cite{ref9}, separates the operator aspect from the time dependence by writing

\begin{equation}
U(t)=\prod_{j}\exp [-i\mu _j(t)A_j],  \label{eqn6}
\end{equation}
where the $A_j$ are operators chosen to be time independent while the $\mu_j(t)$ are classical functions that carry all the time dependence. Our constructive procedure consists of inserting Eq.~(\ref{eqn6}) into Eq.~(\ref{eqn1}) whereupon the derivative on the left-hand side can be rearranged to have the same structure as the right-hand side through repeated application of the Baker-Campbell-Hausdorff (BCH) identity \cite{ref10} and choosing the $\mu_j(t)$ suitably. These functions are seen to satisfy a set of well-defined first-order differential equations. All the operator aspects are handled analytically in the BCH manipulations while the time dependence aspect is isolated into the equations for the $\mu(t)$. These may require numerical solution but that is readily carried out through a simple MATHEMATICA program \cite{ref11}. The same procedure also applies to the Lindblad structure of Eq.~(\ref{eqn5}) after first recasting it into the same form as Eq.~(\ref{eqn1}), namely,

\begin{equation}
i\dot{\eta}(t)=\mathcal{L}(t) \eta(t),
\label{eqn7}
\end{equation}
where $\eta(t)$ are suitable linear combinations, $(n^2-1)$ in number, of the density matrix elements themselves. Such a translation of Eq.~(\ref{eqn5}) into Eq.~(\ref{eqn7}) passes, of course, from the $n \times n$ matrix structure of $H$ and $L_k$ into $(n^2-1) \times (n^2-1)$ matrices for the operators in $\mathcal{L}$. Thus, the Lindblad equation for a two-level system is also of the $3 \times 3$ form of Eq.~(\ref{eqn1}) and Eq.~(\ref{eqn7}) \cite{ref6}.

\section{General solution of a three-level problem}

The above construction requires that all the operators that arise as a result of the BCH application be included in the set of operators $A_j$ in Eq.~(\ref{eqn6}). Clearly, for the most general three-level problem, this requires eight linearly independent matrices which, along with the unit matrix, provide a complete description. The eight traceless matrices used in a standard description of SU(3) \cite{ref12} are a convenient choice. In our previous work \cite{ref6,ref8}, we restricted our attention to forms of $H$ and $L_k$ for which a smaller subset, namely three, sufficed:

\begin{eqnarray}
A_x= \lambda_6 = \left( 
\begin{array}{ccc}
0 & 0 & 0 \\ 
0 & 0 & 1 \\ 
0 & 1 & 0
\end{array}
\right)&,& A_y= \lambda_5 = \left( 
\begin{array}{ccc}
0 & 0 & -i \\ 
0 & 0 & 0 \\ 
i & 0 & 0
\end{array}
\right),\, \nonumber \\
A_z = \lambda_1 &=& \left( 
\begin{array}{ccc}
0 & 1 & 0 \\ 
1 & 0 & 0 \\ 
0 & 0 & 0
\end{array}
\right). 
\label{eqn8}
\end{eqnarray}
This set of three, displaying also their $\lambda$ notation in the SU(3) literature \cite{ref12}, coincide with the SO(3) angular momentum operators of three-dimensional rotations and close under mutual commutation between them. Thus, only three such terms and the unit operator are necessary in Eq.~(\ref{eqn6}), affording a considerable simplification. (It helps to replace $A_x$ and $A_y$ by their linear combinations $A_x \pm iA_y$.) In particular, the resulting set of equations for the three $\mu$ are readily amenable to solution \cite{ref11}, consisting of a Riccati equation (first-order and quadratically nonlinear) for one of them whose solution then leads to simple quadrature solutions for the other two. The $H$ and $L_k$ considered in \cite{ref6,ref8} permitted such a use of the SO(3) sub-group of SU(3) to simplify the solution of Eqs.~(\ref{eqn1}) and (\ref{eqn5}).

However, Hamiltonians such as those in Eqs.~(\ref{eqn2})-(\ref{eqn4}) cannot be expressed in terms of just the three operators in Eq.~(\ref{eqn8}). Similarly, a two-level Lindblad equation \cite{ref12} with
$H=\frac{1}{2}\epsilon (t)\sigma _z+J(t) \sigma _x ,\; L=\sqrt{\Gamma } \sigma _z $, gives an Eq.~(\ref{eqn7}) of the form \cite{ref6}

\begin{eqnarray}
i\frac {d}{dt}
\left( 
\begin{array}{c}
\rho _{12}+\rho _{21} \\ 
\rho _{21}-\rho _{12} \\ 
\rho _{11}-\rho _{22}
\end{array}
\right) & = & \left( 
\begin{array}{ccc}
E_1 & -\epsilon (t) & 0 \\ 
-\epsilon (t) & E_2 & 2J(t) \\ 
0 & 2J(t) & 0
\end{array}
\right)  \nonumber \\ 
& &\quad \quad \times \left( 
\begin{array}{c}
\rho _{12}+\rho _{21} \\ 
\rho _{21}-\rho _{12} \\ 
\rho _{11}-\rho _{22}
\end{array}
\right),
\label{eqn9}
\end{eqnarray}
with $E_1=E_2=-i\Gamma$.
Again, the $\mathcal{L}$ operator requires use of the full set of eight operators of the SU(3) algebra.

Such a set of eight includes, besides the three in Eq.~(\ref{eqn8}), three other off-diagonal matrices \cite{ref12},

\begin{eqnarray}
\lambda_2 = \left( 
\begin{array}{ccc}
0 & -i & 0 \\ 
i & 0 & 0 \\ 
0 & 0 & 0
\end{array}
\right)&,& \lambda_4 = \left( 
\begin{array}{ccc}
0 & 0 & 1 \\ 
0 & 0 & 0 \\ 
1 & 0 & 0
\end{array}
\right),\, \nonumber \\
\lambda_7 &=& \left( 
\begin{array}{ccc}
0 & 0 & 0 \\ 
0 & 0 & -i \\ 
0 & i & 0
\end{array}
\right), 
\label{eqn10}
\end{eqnarray}
and two diagonal ones,

\begin{equation}
\lambda_3 = \left( 
\begin{array}{ccc}
1 & 0 & 0 \\ 
0 & -1 & 0 \\ 
0 & 0 & 0
\end{array}
\right),\,\,\, \lambda_8 = \frac{1}{\sqrt{3}} \left( 
\begin{array}{ccc}
1 & 0 & 0 \\ 
0 & 1 & 0 \\ 
0 & 0 & -2
\end{array}
\right).
\label{eqn11}
\end{equation}
           
All eight operators and the unit operator are needed in Eq.~(\ref{eqn6}) for the general solution. Once again, the linear combinations that simplify the commutators and, therefore, the final set of equations for $\mu$ are $a_{\pm} =\frac{1}{2}(\lambda_6 \pm i\lambda_7), b_{\pm} =\frac{1}{2} (\lambda_1 \mp i\lambda_2), c_{\pm} = \frac{1}{2} (\lambda_4 \pm i\lambda_5)$, and $a_3=\frac{1}{2} (\sqrt{3}\lambda_8 - \lambda_3), c_3= \frac{1}{2} (\sqrt{3}\lambda_8 + \lambda_3)$. Table 1 shows the commutators between this set of eight operators. The three sets of triplets \{a, b, c\} coincide to within factors of $\frac{1}{2}$ with the sets \{V, T, U\} used in the SU(3) literature \cite{ref12}.

\begin{table*} 
\begin{center}
\begin{tabular}{|c||c||c||c||c||c||c||c||c|}

\hline
 $-$&$a_3$&$a_{+}$&$a_{-}$&$c_3$&$c_{+}$&$c_{-}$&$b_{+}$&$b_{-}$  \\ \hline
 \hline
$a_3$&$0$&$2a_{+}$&$-2a_{-}$&$0$&$c_{+}$&$-c_{-}$&$b_{+}$&$-b_{-}$  \\
\hline
$a_{+}$&$-2a_{+}$&$0$&$a_3$&$-a_{+}$&$0$&$b_{+}$&$0$&$-c_{+}$  \\
\hline
$a_{-}$&$2a_{-}$&$-a_3$&$0$&$a_{-}$&$-b_{-}$&$0$&$c_{-}$&$0$  \\
\hline
$c_3$&$0$&$a_{+}$&$-a_{-}$&$0$&$2c_{+}$&$-2c_{-}$&$-b_{+}$&$b_{-}$ \\
\hline
$c_{+}$&$-c_{+}$&$0$&$b_{-}$&$-2c_{+}$&$0$&$c_3$&$-a_{+}$&$0$  \\
\hline
$c_{-}$&$c_{-}$&$-b_{+}$&$0$&$2c_{-}$&$-c_3$&$0$&$0$&$a_{-}$  \\
\hline
$b_{+}$&$-b_{+}$&$0$&$-c_{-}$&$b_{+}$&$a_{+}$&$0$&$0$&$a_3-c_3$  \\
\hline
$b_{-}$&$b_{-}$&$c_{+}$&$0$&$-b_{-}$&$0$&$-a_{-}$&$c_3-a_3$&$0$ \\
\hline
\end{tabular}
\end{center}
\caption{Table of commutators. With operators $O_i$ in the first column and $O_j$ in the top row, each entry provides the commutator $[O_i,O_j]$.}
\end{table*}

Flexibility in the order of the various factors in Eq.~(\ref{eqn6}) is a feature of our technique. Inspection of Table 1 permits optimal ordering such that each application of the BCH identity generates at most two terms. We choose, thereby, the order

\begin{eqnarray}
U(t) & = & e^{-i\delta} e^{-i\mu_8 b_{+}} e^{-i\mu_7 b_{-}} e^{-i\mu_6 c_{+}} e^{-i\mu_5 c_{-}} \nonumber \\
 & & \times \,\, e^{-i\mu_3 a_{+}} e^{-i\mu_2 a_{-}} e^{-i\mu_1 a_3} e^{-i\mu_4 c_3}.
\label{eqn12}
\end{eqnarray}

Evaluating $idU/dt$, re-arranging through use of the BCH identity into the form of an operator sum acting from the left on $U(t)$, the coefficients of $b_{+}, b_{-}, c_{+}, c_{-}, a_{+}, a_{-}, a_3, c_3$ in the operator sum are, respectively, 

\begin{eqnarray}
\!\!\!\!\!\!\!\!\!\!\!\! & &\!\!\!\!\!\!\!\!\!\!\!\!\!\!\! \dot{\mu}_8 +i\mu_5 s +\mu_8 ^2 w +i\mu_8 (-i\mu_3 r +\dot{\mu}_1 -\!\! 2\mu_5 \mu_7 s -\!\! \dot{\mu}_4 +i\mu_6 v) \nonumber \\
w & \equiv & \!\! \dot{\mu}_7 -i\mu_6 r +i\mu_5 \mu_7 ^2 s -\!\! i\mu_7 (-i\mu_3 r +\dot{\mu}_1 +i\mu_6 v -\!\! \dot{\mu}_4 ) \nonumber \\
u & \equiv &\!\! \dot{\mu}_6 +\mu_6 ^2 v +i\mu_6 (-i\mu_3 r +\dot{\mu}_1 +2\dot{\mu}_4 )-\!\! i\mu_7 (1-\!\! \mu_5 \mu_6) s \nonumber \\
 & & v +i\mu_8 (r +i\mu_7 v) \nonumber \\
 & & -i\mu_8 u +(1-\mu_5 \mu_6 ) s \nonumber \\
 & & r +i\mu_7 v \nonumber \\
 & & \dot{\mu}_1 -i\mu_8 w -i\mu_3 r -\mu_5 \mu_7 s \nonumber \\
 & & \dot{\mu}_4 +i\mu_8 w -i\mu_6 v + \mu_5 \mu_7 s,
\label{eqn13}
\end{eqnarray}
along with $\dot{\delta}$ multiplying $\mathcal{I}$. We have defined for convenience,

\begin{eqnarray}
r & \equiv & \dot{\mu}_2 -i\mu_2 (\dot{\mu}_4 +2\dot{\mu}_1), \nonumber \\
s & \equiv & \dot{\mu}_3 + \mu_3 ^2 r +i\mu_3 (\dot{\mu}_4 +2\dot{\mu}_1 ) \nonumber \\
v & \equiv & \dot{\mu}_5 -i\mu_5 (-i\mu_3 r +\dot{\mu}_1 +2\dot{\mu}_4 ).
\label{eqn14}
\end{eqnarray}
The above set of expressions can be matched to any $H$ in Eq.~(\ref{eqn1}) or $\mathcal{L}$ in Eq.~(\ref{eqn7}) with nine arbitrary time-dependent entries to provide defining equations for the $\mu$ and $\delta$ in Eq.~(\ref{eqn12}). This completes the general solution. 

Thus, for the matrix in Eq.~(\ref{eqn9}), which is a linear combination of ($a_{\pm}, b_{\pm}, a_3, c_3$), we have

\begin{eqnarray}
\dot{\mu}_8 +\mu_8 ^2 \epsilon +i\mu_8 (E_2 \!\!-\! E_1)  =\!\!  -\epsilon \!\!-i\mu_5 m, & & \nonumber \\
\dot{\mu}_7 -\!\! i\mu_5 \mu_7^2 m -\!\!2\mu_7 \mu_8 \epsilon  -  i\!\!\mu_7 (E_2-E_1) -\!\!i\mu_6 n  =\!\!  -\epsilon , &  &  \nonumber \\
\dot{\mu}_6 +i\mu_6 \mu_8 (2 \mu_6 J+i \epsilon)\!\! -\!\! i \mu_5 \mu_6 \mu_7 m \!\!+i\mu_6 E_1   = 2i\mu_7 J,  &  & \nonumber \\
\dot{\mu}_5 +i\mu_5 ^2 \mu_7 m -\!\! i\mu_5 \mu_8 (4 \mu_6 J +i \epsilon) -\!\! i \mu_5 E_1  =   -2i \mu_8 J, & &   \nonumber \\ 
\dot{\mu}_4  = \!\! 2\mu_6 \mu_8 J+i\mu_8 \epsilon \!\! -\!\! \mu_5 \mu_7 m  +\frac{1}{3}  (2E_1\!-\!\!E_2),  & & \nonumber \\
\dot{\mu}_3 \!\! -\!\! \mu_3 ^2 n +i\mu_3 (\mu_5 \mu_7 m+\mu_8 [2 \mu_6 J\!\! -\!\!i \epsilon]  +  E_2)  =  m, & &  \nonumber \\
\dot{\mu}_2 \!\!-\!\! i\mu_2 \mu_5 \mu_7 m-\!\! i\mu_2 \mu_8 (2 \mu_6 J-\!\! i \epsilon)+2\mu_2 \mu_3 n  - \!\! i\mu_2 E_2  =   n, & & \nonumber \\
\dot{\mu}_1  \!\!=  i\mu_3 n+\!\!\mu_5 \mu_7 m\!\! -\!\! i\mu_8 \epsilon  + \!\! \frac{1}{3}(2E_2\!\!-\!E_1),   \,\,\,\,\, & & \label{eqn15}
\end{eqnarray}
with $\dot{\delta} =  (E_1+E_2)/3$, where we have defined $m \equiv 2J/(1-\mu_5 \mu_6), n \equiv 2J(1-\mu_7 \mu_8)$.
The above eight equations fall into two groups, the first four involving only $\mu_5 -\mu_8$. They may be solved first and then serve as inputs for solving the remaining four equations. This structure was anticipated in our previous work \cite{ref8}, because the $8 \times 8$ matrices in Eq.~(\ref{eqn7}) that follow from the operators ($a_{\pm}, b_{\pm}, a_3, c_3$) are in $4 \times 4$ block-diagonal form. The above set of equations are, of course, more complicated than in \cite{ref6,ref8}, being more nonlinear but note again the characteristic structure of no higher power than quadratic of any $\mu$. 

After completion of our work, we became aware of closely related papers that use the Wei-Norman \cite{ref9} procedure for three-level problems \cite{ref14,ref15}. In particular, \cite{ref15} uses a product of nine exponentials involving the operators \{T, U, V\} of SU(3) with an arrangement slightly different in order from ours in Eq.~(\ref{eqn12}). Our choice seems to yield a less complicated set of equations in Eq.~(\ref{eqn15}) that define the $\mu$ functions. 

\section{Lindblad equation for a two-level system}

The most general two-level problem with an arbitrary time-dependent Hamiltonian and any choice for the $L_k$ in the master equation in Eq.~(\ref{eqn5}) leads to a $3 \times 3$ equation for the density matrix elements such as Eq.~(\ref{eqn9}) and is thereby solved through the set in Eq.~(\ref{eqn15}). Fig.\ 1 provides an illustration for particular choices of the parameters in Eq.~(\ref{eqn9}). The left-hand side of the panels show perfect agreement with the solutions obtained by a different numerical procedure in \cite{ref13}. The right-hand side shows for comparison our earlier results \cite{ref6} where the decoherence part involving $\Gamma$ was simplified so as to reduce the second term on the right-hand side of Eq.~(\ref{eqn5}) to a unit-diagonal contribution on the right-hand sides of Eq.~(\ref{eqn7}) and Eq.~(\ref{eqn9}). As expected, the difference between the two lies mainly in the off-diagonal density matrix element and the rate of damping. The entropy, while still rising monotonically to $\ln 2$ as the system evolves from a pure state to a mixed one, also differs in these two models of the decoherence. The comparison suggests that the simplified model for decoherence introduced in \cite{ref6} works reasonably well. Since it is much easier to implement, especially for higher $n$, requiring only $n$-dimensional matrices rather than $n^2-1$, we will so use it below for three-level systems.

\begin{figure}
\includegraphics[width=3in]{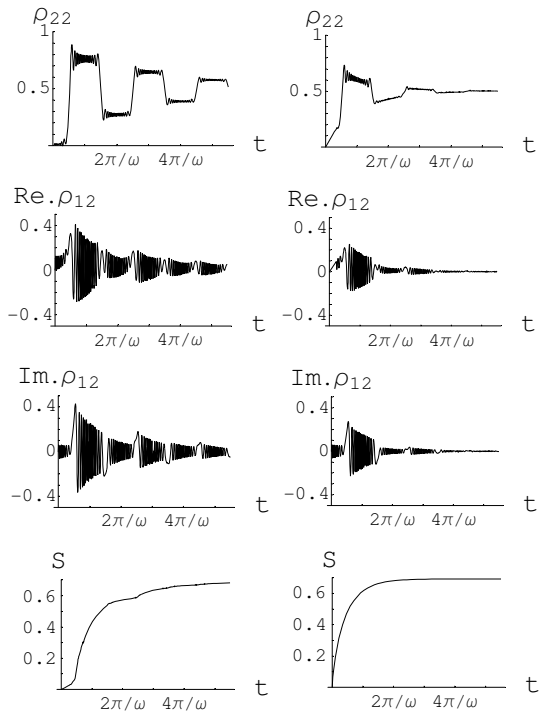}
\caption{Diagonal and off-diagonal elements of the density matrix and entropy $S$ for a two-level system described by Eq.~(\ref{eqn5}) and  Eq.~(\ref{eqn9}). With $\epsilon (t) =A \cos (\omega t), 2J = B \cos (\Omega t + \delta)$, results are shown in the left-hand panels for an initial pure state 1, and $A=45, B=6, \omega =1, \Omega =0, \delta =0, \Gamma = 0.3$. These results coincide with those of \cite{ref13}. The right-hand panels show for comparison the results of a simplified model for decoherence in \cite{ref6}.}
\end{figure}

\section{Applications to three-level systems} 

In a study of population trapping and dispersion in three-level systems, \cite{ref1} considered the Hamiltonian in Eq.~(\ref{eqn2}) for both ``lambda" and ``ladder" systems with $\Delta_1=\omega_2-\omega_1-\omega_a, \Delta_2=\omega_3-\omega_2 \pm \omega_b$, the signs corresponding to lambda and ladder arrangements, respectively, of the three levels $1,2$, and $3$. Solutions were developed for special cases such as equal amplitudes of the driving fields and a phenomenologically introduced damping through a negative imaginary piece in $\omega_2$. Fig.\ 2 presents our results through Eq.~(\ref{eqn5}) and Eq.~(\ref{eqn15}) with the decay introduced as per the model discussed above. In this model, Eq.~(\ref{eqn5}) reduces to Eq.~(\ref{eqn7}) with the decay occurring as $-i\Gamma$ times the unit operator in $\mathcal {L}$. The results coincide with those of \cite{ref1}. But our procedure extends readily to arbitrary values of the parameters in the Hamiltonian and can be applied also to varying amplitudes, frequencies, and phases of the two driving fields. The same numerical solutions of Eq.~(\ref{eqn15}), when inserted into Eq.~(\ref{eqn12}) provide a full solution of the time evolution of all density matrix elements, whatever the time dependences and values of the parameters in the Hamiltonian in Eq.~(\ref{eqn2}) . Note in the bottom panel of Fig.\ 2 that the intermediate level 2 has negligible population throughout, the population flopping back and forth between the extreme levels.

\begin{figure}
\includegraphics[width=3in]{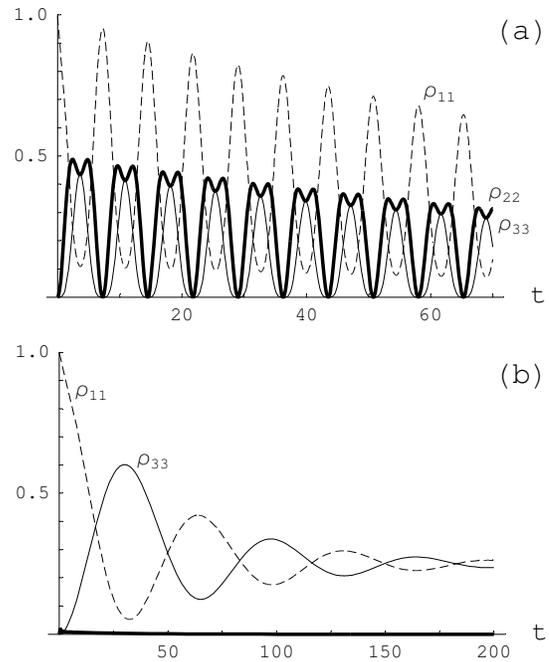}
\caption{Populations of a three-level system described by Eq.~(\ref{eqn2}), starting with initial state 1. The parameters are $|G_1| =|G_2| =0.5$ and (a) $\Delta_1 =  0.5 -0.01 i, \Delta_2 = 0.5 +0.01i$, (b) $\Delta_1 = -\Delta_2 = 5 - i$. The intermediate level 2, shown by a thick line, remains essentially at zero for all time in (b). The results coincide with those of \cite{ref1}.}
\end{figure}

Another more recent paper \cite{ref3} presented an analytical model of three-level systems with the Hamiltonian in Eq.~(\ref{eqn4}) but again under restrictive conditions where the two driving fields have the same hyperbolic-secant time dependence in $\sigma t $, differing only in amplitude, and with the real part of the detuning $\Delta$ set equal to zero. The value of $\sigma $ sets the time scale. In Fig.\ 3 we present our results for a wider range of parameters. Our results for the diagonal components (the off-diagonal ones are not displayed but are also available in our calculations) of the density matrix are presented as functions of time but are equivalent to the display in terms of detuning given in \cite{ref3}. The change in form with increasing values of the parameters $\alpha=\sqrt {A_{12}^2 + A_{23} ^2} \sigma $, $\delta = \frac{1}{2} \Delta \sigma $, and $\gamma = \frac{1}{2} \Gamma \sigma $ are interesting. The population of states 1 and 3 at large $t$ depend critically on these parameters. Here the $A$ are the amplitudes of the two driving fields, and $\Gamma$ the damping (with $-i\Gamma$ added to $\Delta$ in Eq.~(\ref{eqn4})) as in \cite {ref3}. Our results provide a method for exploring a broad range of parameter values and time dependences in Eq.~(\ref{eqn4}).

\begin{figure}
\includegraphics[width=3in]{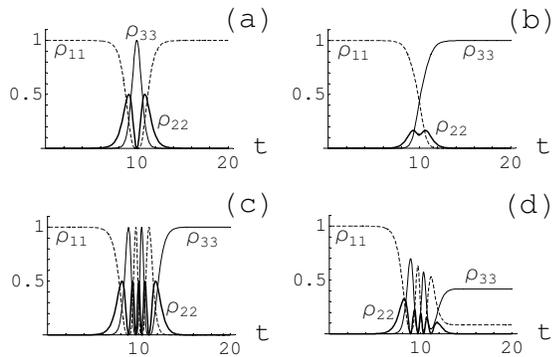}
\caption{Populations of a three-level system described by Eq.~(\ref{eqn4}), starting with initial state 1. Thick and thin lines describe levels 2 and 3, respectively, and a dashed line the population of level 1. Values of parameters described in the text are (a) $\alpha =2, \gamma =0, \delta=0$, (b) $\alpha =2, \gamma =0, \delta=0.866$, (c) $\alpha =5, \gamma =0, \delta=0$, (d) $\alpha =5, \gamma =0.5, \delta=0$. }
\end{figure}

\begin{figure}[h]
\includegraphics[width=3in]{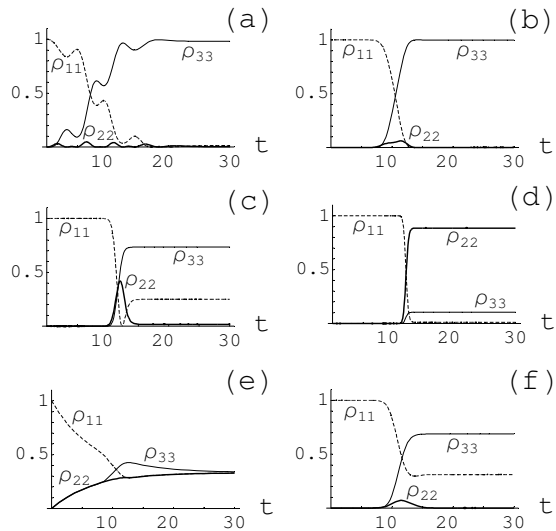}
\caption{Populations of a three-level system described by Eq.~(\ref{eqn4}) , starting with initial state 1. Values of parameters described in the text are $A_2=A_1=2.5, t_1=12, t_2=t_1-\sigma $, and (a) $ \sigma =8$, (b) $\sigma = 3$, (c) $\sigma =1.5$, (d) $\sigma  =0.9$, all with $\gamma =0$ and $\Delta=0$, (e) $\sigma = 3, \gamma= 0.15, \Delta=0 $, (f) $\sigma = 3, \gamma = 0, \Delta = 1.5$. }
\end{figure}

Finally, as yet another application, we consider a very recent paper \cite{ref4} that used a Hamiltonian encoding scheme based on computational control techniques for handling the Dyson series in time-dependent evolution. The results presented were for zero detuning $\Delta$ in which case the problem in Eq.~(\ref{eqn4}) actually reduces to the much simpler $3 \times 3$ problem we considered previously \cite{ref8}. But, we present in Fig.\ 4 much more general results for a wide range of parameters in that Hamiltonian. Decay is again introduced through the simplified model (with $-i\Gamma $ added to the diagonal entries in Eq.~(\ref{eqn4})) that permits application of the $8 \times 8$ problem in Eq.~(\ref{eqn7}) and Eq.~(\ref{eqn15}). The form of the driving fields in Eq.~(\ref{eqn4}) is $\Omega_{12}=(A_1/2) \exp (-[(t-t_1)/\sigma]^2)$ and a similar expression for $\Omega_{23}$ with subscript 2 for the parameters. Starting with initial population in the state 1, Fig.\ 4 shows the subsequent evolution of the three populations. Larger values of $\sigma$ show initial oscillations as the population is transferred from 1 to 3 asymptotically. The effects of damping when all three populations equalize, as well as the effect of increasing $\Delta$ are also shown.  

\section{Summary}

We have developed a complete solution of the time evolution of three-level systems. With individual, arbitrary time-dependent entries in the $3 \times 3$ Hamiltonian, equating those entries to the set in Eq.~(\ref{eqn13}) provides coupled, first-order differential equations for classical functions $\mu$. They are amenable to simple integration through \cite{ref11} and the solutions, when inserted into Eq.~(\ref{eqn12}), provide a complete solution of the quantal problem. Thereby, there is no need to invoke any familiar approximation schemes used such as the rotating-wave approximation. Master equations for two-level systems with dissipation and decoherence, when the most general form is of $3 \times 3$ matrix form, are also solved in the same manner. A simplified model of decoherence allows solution of three-level problems as well. Illustrations are given of a variety of applications from the recent literature for population transfer and trapping in quantum optics and for multiphoton transitions in molecules.

This work has been supported by the U.S. Department of Energy under Grant No. DE-FG02-02ER46018.


\begin{thebibliography}{}

\bibitem[*]{} Email: arau@phys.lsu.edu
 
\bibitem{ref1} P.\ M.\ Radmore and P.\ L.\ Knight, J.\ Phys.\ B {\bf 15}, 561 (1982).
 
\bibitem{ref2} G.\ N.\ Gibson, Phys.\  Rev.\ A {\bf 67}, 042322 (2003).

\bibitem{ref3} N.\ V.\ Vitanov, J.\ Phys.\ B {\bf 31}, 709 (1998).
 
\bibitem{ref4} A.\ Mitra, I.\ R.\ Sola, and H.\ Rabitz, Phys.\ Rev.\ A {\bf 67}, 043409 (2003).
 
\bibitem{ref5} G.\ Lindblad, Commun. Math. Phys. {\bf 48}, 119 (1976); V.\ Gorini, A.\ Kassokowski, and E.\ C.\ G.\ Sudarshan, J. Math. Phys. {\bf 17}, 821 (1976);  D. F.\ Walls and G.\ J.\ Milburn, \textit{Quantum Optics} (Springer-Verlag, Berlin, 1994); M.\ O.\ Scully and M.\ S.\ 
Zubairy, \textit{Quantum Optics} (Cambridge Univ. Pr., 1996); W.\ P.\ Schleich, 
\textit{Quantum Optics in Phase Space} (Wiley-VCH, Berlin, 2001); D.\ Giulini, E.\ Joos, C.\ Kiefer, J. \ Kupsch, I.\ O.\ Stamatescu, and H.\ D.\ Zeh, \textit{Decoherence and the Appearance of a Classical World in Quantum Theory} (Springer-Verlag, Berlin, 1996).

\bibitem{ref6} A.\ R.\ P.\ Rau and R.\ A.\ Wendell, Phys. Rev. Lett. {\bf 89}, 220405(1-4) (2002).

\bibitem{ref7} See, for instance, R.\ Alicki and K.\ Lendi, \textit{Quantum Dynamical Semigroups and Applications} (Springer-Verlag, Berlin, 1987); E.\ B.\ Davies, \textit{Quantum Theory of Open Systems} (Academic Press, London, 1976).

\bibitem{ref8} A.\ R.\ P.\ Rau and Weichang Zhao, Phys.\ Rev.\ A {\bf 68}, 052102(1-6) (2003).

\bibitem{ref9} J.\ Wei and E.\ Norman, J.\ Math.\ Phys.\ {\bf 4}, 575 (1963).

\bibitem{ref10} See, for instance, J.\ J.\ Sakurai, \textit{Modern Quantum Mechanics} (Addison-Wesley, Reading, MA, 1994), Sec.\  2.3.

\bibitem{ref11} S.\ Wolfram, {\em Mathematica: A System for Doing Mathematics by Computer} (Addison-Wesley, Redwood City, CA, 1988).

\bibitem{ref12} See, for instance, A.\ W.\ Joshi, \textit{Elements of group theory for physicists}, p.145

\bibitem{ref13} Y.\ Kayanuma, Phys.\ Rev.\ B {\bf 47}, 9940 (1993); Y.\ Kayanuma and Y.\ Mizumoto, Phys.\ Rev.\ A {\bf 62}, 061401 (2000); K.\ Saito and Y.\ Kayanuma, Phys.\ Rev.\ A {\bf 65}, 033407 (2002).

\bibitem{ref14} D.\ V.\ Fursa and G.\ L.\ Yudin, Phys.\ Rev.\ A {\bf 44}, 7414 (1991). We thank Dr.\ Fursa for bringing this paper to our attention which also led us to \cite{ref15} below.

\bibitem{ref15} G.\ Dattoli, J.\ C.\ Gallardo, and A.\ Torre, Riv.\ Nuovo Cimento {\bf 11}, No.\ 11, 1 (1988), and references therein.
 
 \end{thebibliography}
\end{document}